# Thermally reconfigurable extraordinary terahertz transmission using vanadium dioxide

**S. HADI BADRI[1], HADI SOOFI[2], SANAM SAEIDNAHAEI[3,*]**

[1] *Department of Electrical Engineering, Sarab Branch, Islamic Azad University, Sarab, Iran.*
[2] *Faculty of Electrical and Computer Engineering, University of Tabriz, Tabriz, Iran.*
[3] *Department of Physics, Yeungnam University, Gyeongsan, 38541, Republic of Korea.*
[*] *sanam.nahaei@yu.ac.kr*

**Abstract:** We numerically demonstrate a reconfigurable extraordinary terahertz transmission based on a phase-change material of vanadium dioxide ($VO_2$). The proposed hybrid metasurface is composed of an array of subwavelength apertures perforated on a gold film. The holes are partially filled with annular $VO_2$ and gold disks to control the effective aperture area and the modes inside the aperture. Switching between the insulator and the metallic phase of $VO_2$ provides a convenient way to shift the transmission window. We present two designs offering redshift or blueshift of the extraordinary terahertz transmission. Upon phase transition from the insulator to the metallic phase, in the first design, the transmission peak redshifts from 1.02 to 0.82 THz while in the second design the transmission peak blueshifts from 0.71 to 0.77 THz. Furthermore, the transmission level and resonance frequency can be modulated by controlling the partial phase transition of the $VO_2$. The potential applications for the proposed structures are terahertz modulators and reconfigurable filters.

## 1. Introduction

The incident field is efficiently reflected from the metal film without any perforation. Bethe's theory predicts that the transmission from a metallic film with a hole is proportionate to $(r/\lambda)^4$ where $r$ is the radius of the hole and $\lambda$ is the wavelength of impinging light. Bethe also predicted that the emerging electromagnetic field would diffract. However, transmission resonances at wavelengths larger than the hole radius have been observed in an array of subwavelength apertures perforated on metallic films with lower-than-expected diffraction. This phenomenon is referred to as extraordinary optical transmission (EOT) [1, 2]. Adding an array of holes to the metal film changes the characteristics of the surface plasmon modes and helps to conserve momentum. Consequently, the incident electromagnetic radiation can be coupled to the leaky surface plasmon modes at both sides of the metal film. The two surface plasmon modes are weakly coupled to a continuum of radiation modes; moreover, the two modes are weakly coupled between themselves through the holes leading to a transmission resonance. The collective response of an array of subwavelength holes determines the transmission spectrum of the structure [2, 3]. Therefore, the geometry factors of the holes such as their symmetry, aspect ratio, shape, and aperture area determine the electromagnetic field distribution of the surface plasmon modes and their scattering efficiencies [1]. The high transmission and local field enhancement of EOT can be achieved at various wavelengths by engineering the subwavelength apertures on the metallic films [3]. These properties of EOT have led to interesting applications such as sensing [4, 5], filters [6, 7], lenses [8, 9], optical rotation [10], metamaterials [11, 12], and enhancement of nonlinear effects [13, 14].

In this paragraph, the main techniques for controlling and tuning the EOT are reviewed, including superconductor, graphene, nematic liquid crystal, Schottky diode, piezoelectric, and phase-change materials. The operating temperature has been controlled to modulate the transmission level of a superconductor with a series of subwavelength holes. When the temperature is raised from 8.2 to 18 K, the transmission peak at 0.59 THz drops from 0.98 to 0.48 [15]. To develop transmission peaks at 1.77 and 2.42 THz, a ring-rod nested structure in a metallic film aperture with graphene placed beneath the metallic rod has been introduced. The transmission level at 2.42 THz drops from 0.7 to 0.06 when the Fermi energy is changed from 0 to 0.5 eV, whereas the transmission level at 1.77 THz increases slightly [16]. Graphene ribbons placed in the subwavelength metallic slits have been utilized to modulate the EOT. The measurements indicate that the modulation efficiency of 28.6% can be achieved in transmission at 1397 $cm^{-1}$ [17]. By controlling the birefringence of the nematic liquid crystal filling one side of the metallic hole array, the peak transmission frequency can be tuned from 0.193 to 0.188 THz while the peak transmission increases from 0.58 to 0.70 [18]. A subwavelength metal hole array fabricated on a doped semiconductor forming a Schottky diode structure has been introduced to control the resonance-enhanced terahertz transmission. Controlling the bias voltage changes the depletion zone and, as a result, the semiconductor's conductivity. Under reverse bias of 0 to 16 volts, the measured transmission peak at 0.75 THz increases from about 0.018 to 0.038, resulting in the intensity modulation depth of 35.7% [19]. Surface acoustic waves have been used to adjust the transmission resonance frequency of a periodically nanostructured metallic film deposited on a piezoelectric material. An acousto-optic variation of $\Delta n$=0.02 in the refractive index shifts the resonance wavelength from 1461.5 to 1474.5 nm [20]. A perforated gold film deposited on a layer of $VO_2$ has been proposed to modulate the transmission level through the structure. The transmission intensity at 2250 nm is modulated from around 0.5 to near 0 when the temperature is increased from 30°C to 80°C [21]. A metasurface consisting of a gold film perforated with an array of circular apertures placed on the phase-change material of $Ge_2Sb_2Te_5$ has been reported. By exciting the continuous $Ge_2Sb_2Te_5$ film by a nanosecond pulsed laser, the transmission level of the metasurface at the frequency of 0.85 THz can be modulated with an efficiency of 88% [22].

Metasurfaces are two-dimensional metamaterials with unique applications [23-26]. Recently, phase-change materials (PCMs) with large changes in their optical properties have been incorporated in designing reconfigurable or active metasurfaces [27-29] and other integrated photonic devices [30-32]. $VO_2$ is a typical PCM with metallic and insulator phases. The transition between these phases can be triggered reversibly by controlling the operating temperature of the device [29]. The difference in electric conductivity between the insulator and metallic phases of $VO_2$ in the THz range is around three orders of magnitude, paving the way to design novel devices. In this paper, we present a reconfigurable metasurface composed of an array of subwavelength holes on a metallic film filled partially with annular $VO_2$ and metallic disks. The phase-transition of $VO_2$ enables us to control the incident terahertz electromagnetic field squeezing through the subwavelength apertures. Some of the earlier studies on tunable EOT designs only offer amplitude modulation of EOT while others support only resonance frequency blueshifting or redshifting. Furthermore, the reported frequency shifts should be improved and larger frequency shifts are desirable. By presenting structures offering higher redshift or blueshift, we hope to overcome these difficulties. We propose two designs; one offers a redshift and the other supports a blueshift of the transmission resonance, upon increasing the operating temperature. In the first design, the outer $VO_2$ and the inner gold disks partially fill the holes. In this design, when the $VO_2$ is in the insulator phase, the transmission peak occurs at near 1.02 THz. Upon phase transition to the metallic phase, the transmission resonance redshifts to 0.82 THz. In the second design, we reverse the position of $VO_2$ and gold disks, such that the inner $VO_2$ and the outer gold disks partially fill the holes. The transition of the $VO_2$ disks from the insulator to the metallic phase blueshifts the transmission peak from 0.71 to 0.77 THz. Moreover, we numerically investigate the effect of geometrical parameters and partial phase transition of $VO_2$ on the EOT of the metasurface. When compared to earlier designs, the developed structures accomplish larger frequency shifts while maintaining a relatively narrow bandwidth. Our developed structures could be used as frequency modulators or reconfigurable filters since they have larger frequency shifts and fewer transmission amplitude changes.

## 2. Designing a redshifting metasurface with EOT

A typical method of achieving extraordinary optical transmission is perforating a metallic film with periodically distributed holes. Here, we partially fill the circular holes with metal and a phase-change material of $VO_2$. The phase transition of $VO_2$ from the insulator to the metallic phase in our first designed metasurface results in a redshift of the EOT. As shown in Fig. 1, the gold film with a thickness much larger than the skin depth ($t_m$=0.2 μm) is placed on the $SiO_2$ substrate with a thickness of $t_s$=75 μm. Holes with a radius of $R_3$=50 μm and a lattice constant of $P$=150 μm are perforated on a gold film. The holes are filled with gold and $VO_2$ annular disks while the sizes of them are determined by $R_1$=25 μm and $R_2$=40 μm. The thickness of the $VO_2$ is the same as the gold film. The full-wave frequency domain solver of the CST Microwave Studio software is utilized to evaluate the performance of the designed metasurfaces. The periodic boundary condition along the x- and y-axes are applied to the unit cell while a plane wave is impinging into the metasurface at normal incidence. We should point out that the structure is symmetric and hence it is insensitive to the polarization of the incident electromagnetic field. The transmission through the structure is calculated as $T(\omega)=|S_{21}|^2$. The relative permittivity of $VO_2$ is described by the Drude model.

$$\varepsilon_{VO_2}(\omega) = \varepsilon_\infty - \frac{\omega_p^2(\sigma)}{\omega^2 + i\gamma\omega} \tag{1}$$

where $\varepsilon_\infty = 12$ is the dielectric permittivity at very high frequencies. The plasma frequency at $\sigma$ is $\omega_p^2(\sigma) = \omega_p^2(\sigma_0)\sigma/\sigma_0$ where $\sigma_0 = 3\times10^5\,\Omega^{-1}cm^{-1}$ and $\omega_p(\sigma_0) = 1.4\times10^{15}\,rad/s$. And the collision frequency is $\gamma = 5.75\times10^{13}\,rad/s$ [33]. At approximately 25°C and 85°C, the $VO_2$ is in the insulator and metallic phases corresponding to $\sigma$ of $2\times10^2$ and $2\times10^5$ S/m, respectively [34]. The optical properties of the gold are also described by the Drude model.

$$\varepsilon_{Au}(\omega) = 1 - \frac{\omega_p^2}{\omega^2 + i\gamma\omega} \tag{2}$$

where the plasma frequency is $\omega_p = 1.37 \times 10^{16} rad/s$ while the collision frequency is $\gamma = 1.2 \times 10^{14} rad/s$ [35]. At the terahertz frequencies, the $SiO_2$ can be considered as a lossless medium with a dielectric constant of 3.8 [35].

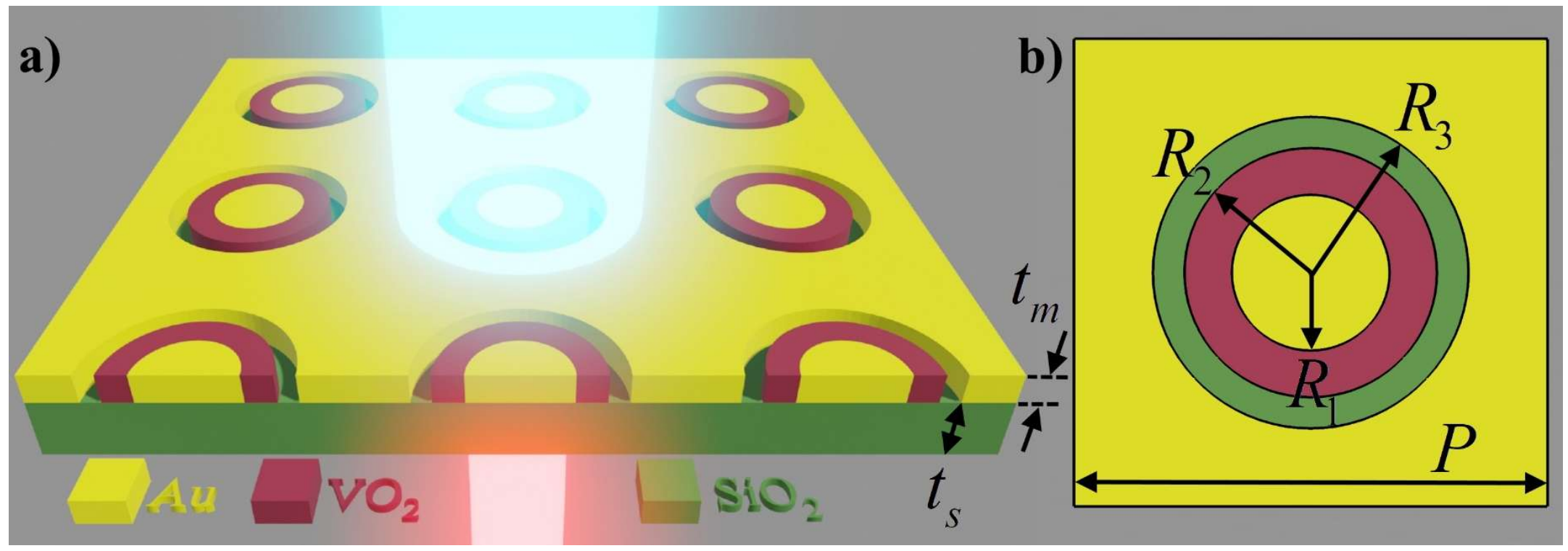

Fig. 1. a) The structure of the tunable metasurface with transmission resonances. b) The top view of the unit cell. The thicknesses of the films and the substrate are not to scale.

The transmission spectra of the designed metasurface for the insulator and the metallic phases of $VO_2$ are shown in Fig. 2. At room temperature, the $VO_2$ is in the insulator phase and the transmission peak of 0.78 occurs at 1.02 THz. When the operating temperature increase to 85°C, the $VO_2$ transitions to the metallic phase and the transmission resonance redshifts to 0.82 THz with an amplitude of 0.67. Therefore, a frequency redshift of 0.2 THz can be achieved upon the transition of the $VO_2$ from the insulator to the metallic phase. The bandwidth in the insulator and metallic phases is approximately 0.13 and 0.12 THz, respectively. We define the transmission bandwidth as a window with transmission levels higher than 0.5. The distribution of the electric field is displayed in Fig. 3 for the insulator and metallic phases of $VO_2$ at their corresponding resonance frequencies. One of the dominant explanations for the EOT is the weak coupling of leaky surface plasmon modes, on the top and the bottom surfaces of the perforated metal film, through the holes [2, 3]. Therefore, the geometry of the periodic array of holes plays a crucial role in boosting the transmission. As shown in Fig. 3, in the insulator phase of $VO_2$, the effective aperture area is larger than the case with the metallic phase. Due to the phase transition of $VO_2$, the effective aperture area and the mode inside the hole change. This manipulates the coupling of the surface plasmon modes through the holes, which in turn, changes the resonance frequency of the metasurface. We discuss the effect of hole geometry on the transmission of the designed metasurfaces in the following subsection. The Cartesian multipole decomposition is also calculated to evaluate the scattering powers of multipoles on the transmission profile of the metasurface. This method has been used to investigate the physical mechanism of resonances in metasurfaces [36, 37]. The multipole decomposition calculations are based on the induced current density ($\boldsymbol{J}(\boldsymbol{r})$) [38]

$$\boldsymbol{J}(\boldsymbol{r}) = -i\omega\varepsilon_0(n^2-1)\boldsymbol{E}(\boldsymbol{r}) \tag{3}$$

where $\boldsymbol{E}(\boldsymbol{r})$ is the electric field distribution and $n$ is the refractive index distribution while $\boldsymbol{r}$ is a position vector from the origin to point $(x, y, z)$, $\varepsilon_0$ is the permittivity of the free space, and $\omega$ is the angular frequency. In the designed metasurface the dominant multipole decomposition terms are the electric dipole ($\boldsymbol{p}$), magnetic dipole ($\boldsymbol{m}$), electric quadrupole ($Q^e$), and magnetic quadrupole ($Q^m$) [39]

$$\boldsymbol{P} = \frac{1}{i\omega}\int \boldsymbol{j} d^3r \tag{4a}$$

$$\boldsymbol{M} = \frac{1}{2c}\int (\boldsymbol{r}\times\boldsymbol{j}) d^3r \tag{4b}$$

$$\boldsymbol{Q}^{e}_{\alpha\beta} = \frac{1}{2i\omega}\int\left[r_\alpha j_\beta + r_\beta j_\alpha - \frac{2}{3}(\boldsymbol{r}\cdot\boldsymbol{j})\delta_{\alpha,\beta}\right]d^3r \tag{4c}$$

$$\boldsymbol{Q}^{m}_{\alpha\beta} = \frac{1}{2c}\int\left[(\boldsymbol{r}\times\boldsymbol{j})_\alpha r_\beta + (\boldsymbol{r}\times\boldsymbol{j})_\beta r_\alpha\right]d^3r \tag{4d}$$

where c is the speed of light and the summation indices ($\alpha$ and $\beta$) run over the Cartesian coordinates ($x$, $y$, and $z$). The effect of other higher-order modes such as the toroidal dipole is negligible on the scattered intensity. Consequently, the total scattering cross-section can be calculated as the following sum [40, 41]

$$\begin{aligned}\sigma^{total}_{sca} &= \sigma^{p}_{sca} + \sigma^{m}_{sca} + \sigma^{Q^e}_{sca} + \sigma^{Q^m}_{sca} + \ldots \\ &= \frac{k^4}{6\pi\varepsilon_0^2\left|\boldsymbol{E}_{inc}\right|^2}\left[\sum_\alpha\left(\left|p_\alpha\right|^2 + \frac{\left|m_\alpha\right|^2}{c}\right) + \frac{1}{120}\sum_{\alpha\beta}\left(\left|kQ^{e}_{\alpha\beta}\right|^2 + \left|\frac{kQ^{m}_{\alpha\beta}}{c}\right|^2\right) + \ldots\right]\end{aligned} \tag{5}$$

where $k$ is the wavenumber and $\boldsymbol{E}_{inc}$ is the amplitude of the incident plane wave. The Lumerical FDTD was utilized to export the electric field and refractive index distribution. Then the induced current density described in Eq. 3 is calculated by MATLAB. The open-source tool introduced in [38] is employed to calculate multipole moments of Eq. 4. A shown in Fig. 4, the total scattering cross-section is defined by the lower multipoles accurately. In the insulator and metallic phases, the electric dipole is dominant. In the Lumerical simulations, the maximum transmission resonances occur at about 0.83 and 0.69 THz corresponding to the insulator and metallic phases, respectively. The difference between the resonance frequencies calculated by CST and Lumerical stems from their differences in the calculation method, meshing, and perfectly matched layer (PML) boundary conditions.

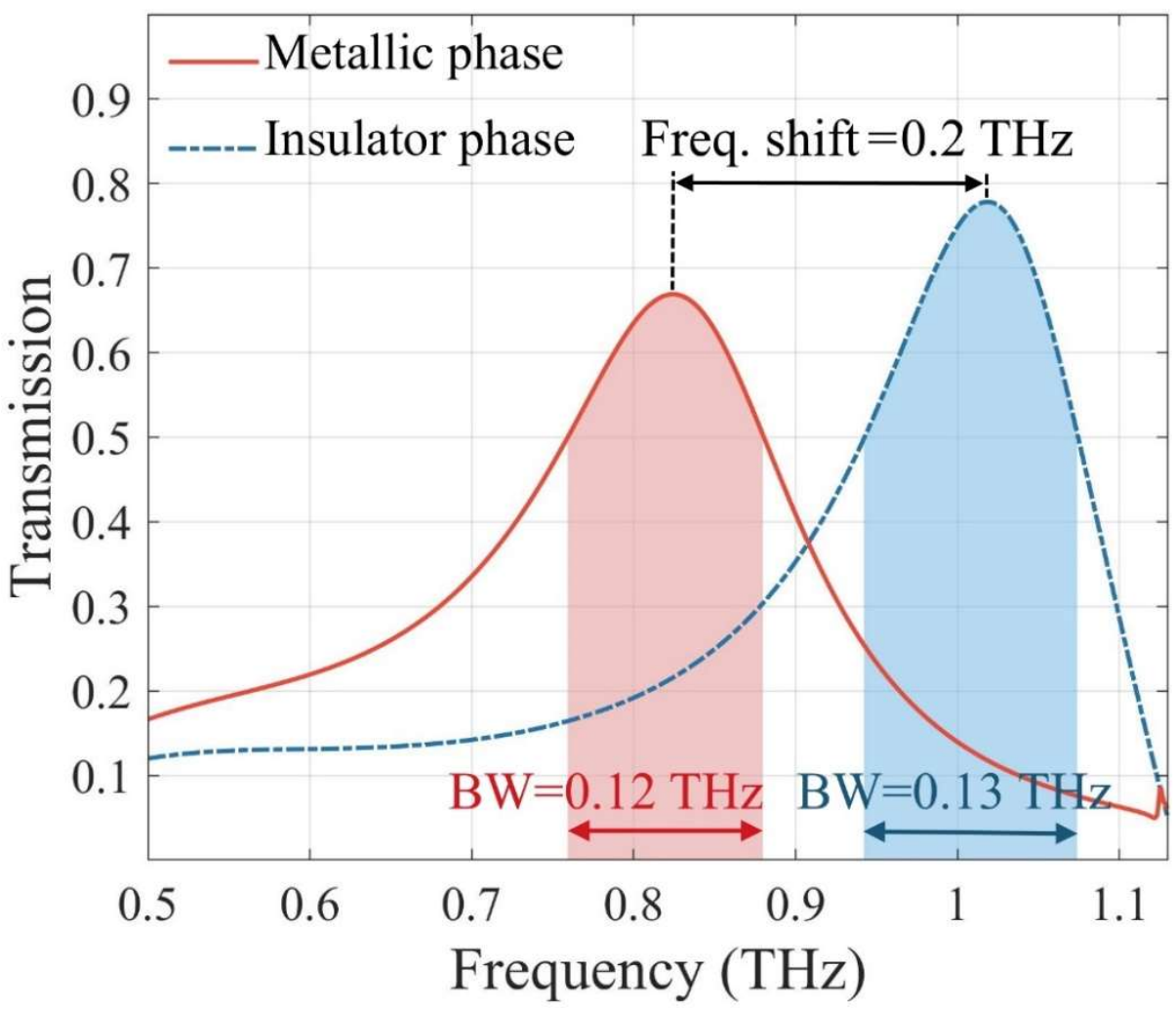

Fig. 2. The transmission spectra of the metasurface for two phases of $VO_2$.

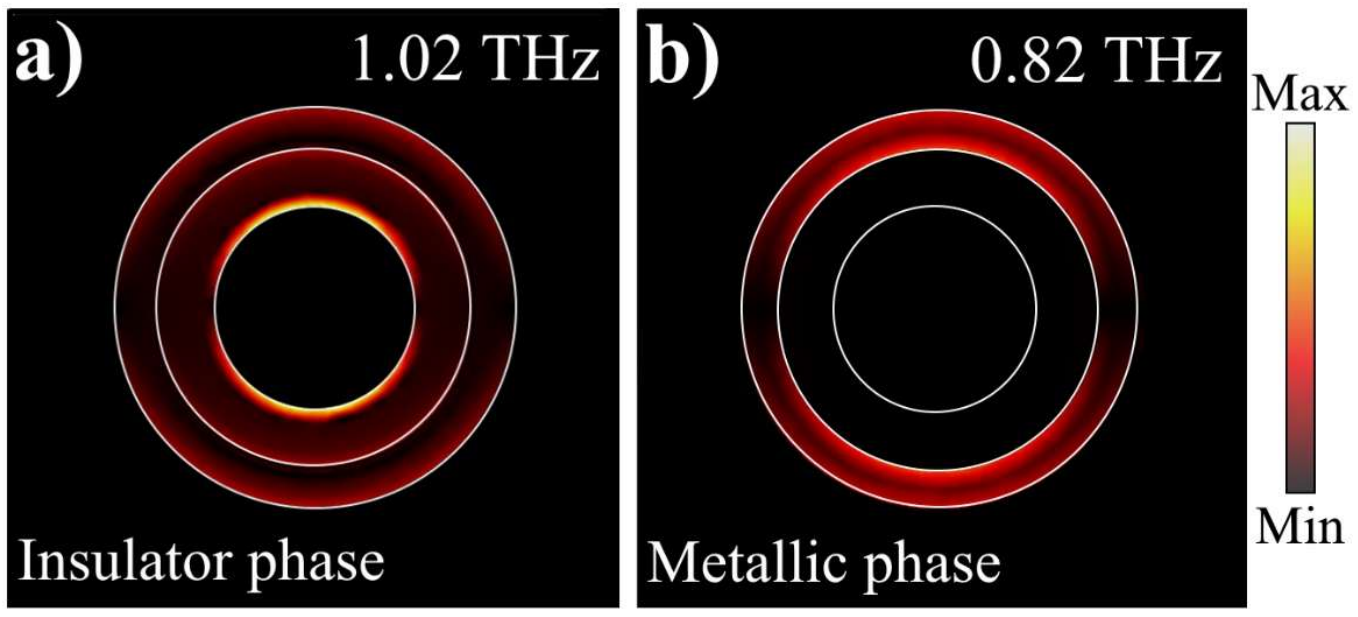

Fig. 3. The electric field distribution in a) the insulator and b) the metallic phases at the transmission resonance of 1.02 and 0.82 THz, respectively.

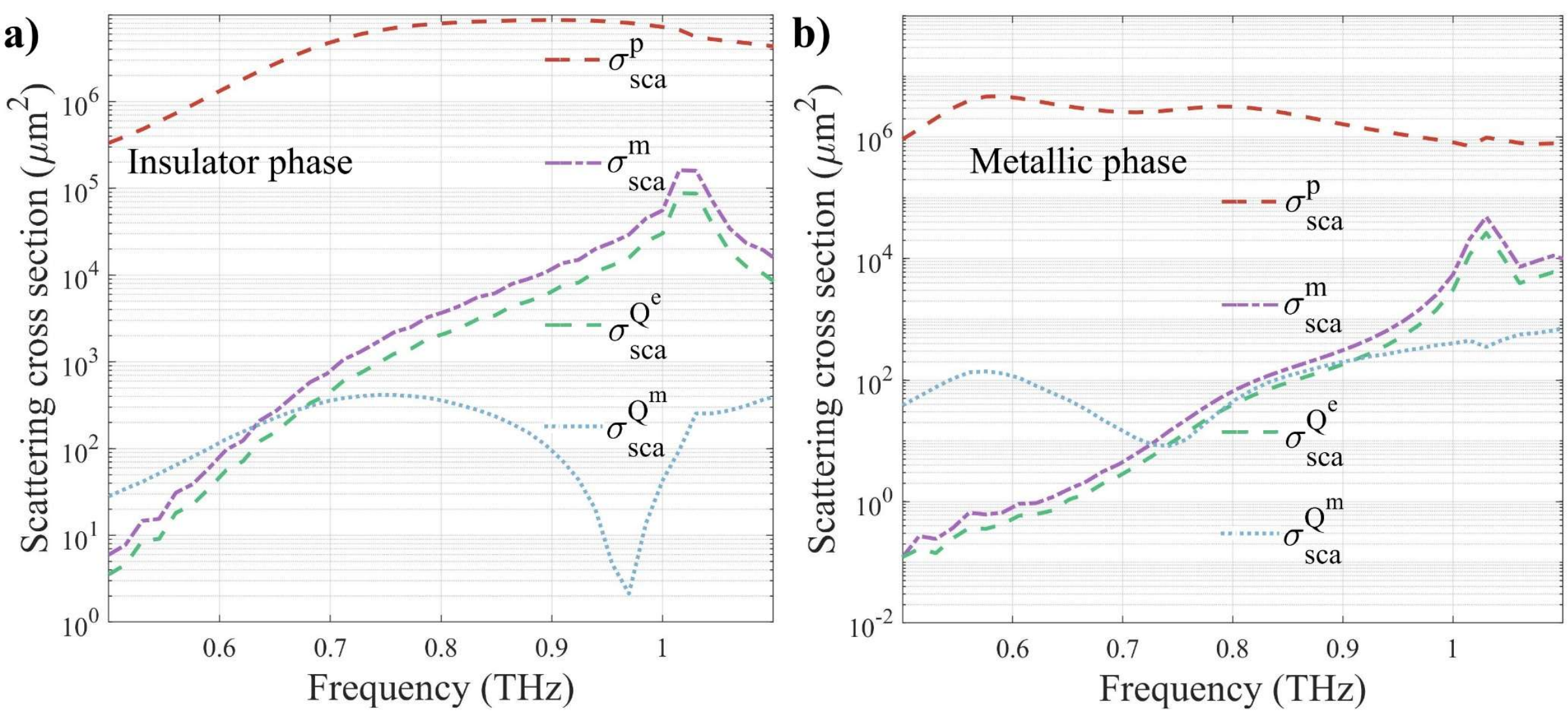

Fig. 4. Scattering cross-sections for a) the insulator and b) the metallic phases associated with the electric dipole (***p***), magnetic dipole (***m***), electric quadrupole ($Q^e$), and magnetic quadrupole ($Q^m$) computed using multipole decomposition.

### *2.1 Effect of hole geometry on EOT*

In this subsection, we examine the effect of the $VO_2$ and the gold disks' size on the transmission resonances of the metasurface while the other geometrical parameters are fixed to $t_m$=0.2 µm, $t_s$=75 µm, $R_3$=50 µm, and $P$=150 µm. First, we examine the effect of $R_1$ (the gold disk's radius) on the transmission resonance while $R_2$ is fixed to 40 µm. The size of the annular $VO_2$ disk is determined by the inner and the outer radii of $R_1$ and $R_2$, respectively. In the metallic phase, the hole can be assumed effectively filled with a disk of radius 40 µm. Hence, as can be seen in Fig. 5, in the metallic phase, changing $R_1$ does not affect the transmission resonance and the resonance occurs at the frequency of 0.82 THz with a transmission peak of 0.67. However, in the insulator phase of $VO_2$, the transmission resonance depends on the radius of the gold disk. For $R_1$=30 µm, the transmission peak of 0.72 occurs at the frequency of 0.97 THz. Reducing the radius of the gold disk to $R_1$=25 µm blueshifts the transmission resonance to 1.02 THz while the transmission peak increases to 0.78. For $R_1$=20 µm, the transmission resonance shifts to 1.04 THz with a maximum transmission level of 0.84. In the case that the hole is only filled with $VO_2$ ($R_1$=0), a transmission peak of 0.91 occurs at 1.06 THz (not shown in Fig. 5). Therefore, the maximum frequency shift of 0.22 THz can be achieved with $R_1$=0 while the difference between the transmission levels increases to 0.24. We chose $R_1$=25 µm where the difference between the transmission levels in the insulator and metallic phases is lower (i.e., 0.11) while the frequency redshift of 0.20 is achieved upon the insulator to the metallic phase transition.

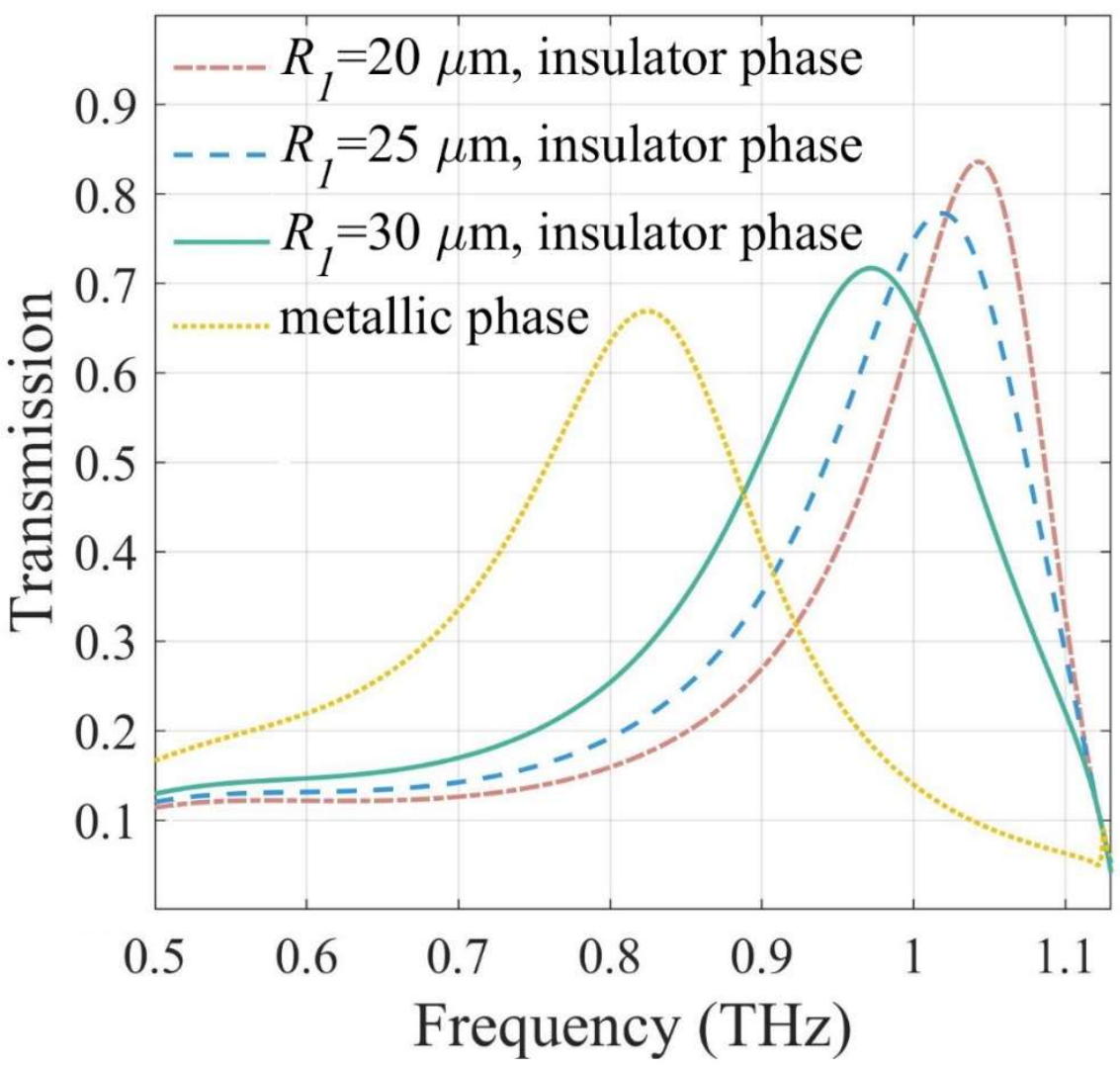

Fig. 5. Transmission resonance vs. the gold disk's radius, with all the other geometrical parameters set to $t_m$=0.2 µm, $t_s$=75 µm, $R_3$=50 µm, $R_2$=40 µm, and $P$=150 µm. In the insulator phase, the transmission resonance redshifts as the radius of the gold disk increases. In the metallic phase, the gold disk's radius has no effect on the transmission resonance since the hole shape is determined by the outer radius of the $VO_2$ annular disk.

Next, we discuss the effect of the annular $VO_2$ disk's size on the transmission resonance. We fix the inner radius of the annular $VO_2$ disk to $R_1$=25 µm. As can be seen in Fig. 6, in the insulator phase, the $VO_2$ disk does not affect the resonance of the structure and the transmission only depends on the gold disk with a fixed radius of $R_1$=25 µm. In the insulator phase of $VO_2$, the transmission resonance occurs at 1.02 THz with a transmission peak of 0.78. Transition to the metallic phase, redshifts the resonance to near 0.90 THz with a transmission peak of 0.78 for $R_2$=35 µm. Increasing the outer radius of the annular $VO_2$ disk to $R_2$=40 µm, redshifts the transmission resonance to 0.82 THz with a maximum transmission level of 0.67. For $R_2$=45 µm, the transmission peak of 0.52 occurs at 0.74 THz, in the metallic phase. In this case, the maximum frequency shift of 0.28 THz is achieved with $R_2$=45 µm, however, the difference between the transmission peaks of the insulator and the metallic phases is 0.26. We chose $R_2$=40 µm with a moderate difference between the transmission peaks of 0.11 and the frequency redshift of 0.20 THz. The simulation results, shown in Fig. 5 and 5, indicate that manipulating the effective aperture area and the mode of the aperture changes the frequency and amplitude of the transmission resonance. We should also point out that the transmission resonances also depend on the other geometrical parameters such as the hole radius ($R_3$), period ($P$), and substrate thickness ($t_s$). However, we do not discuss them here.

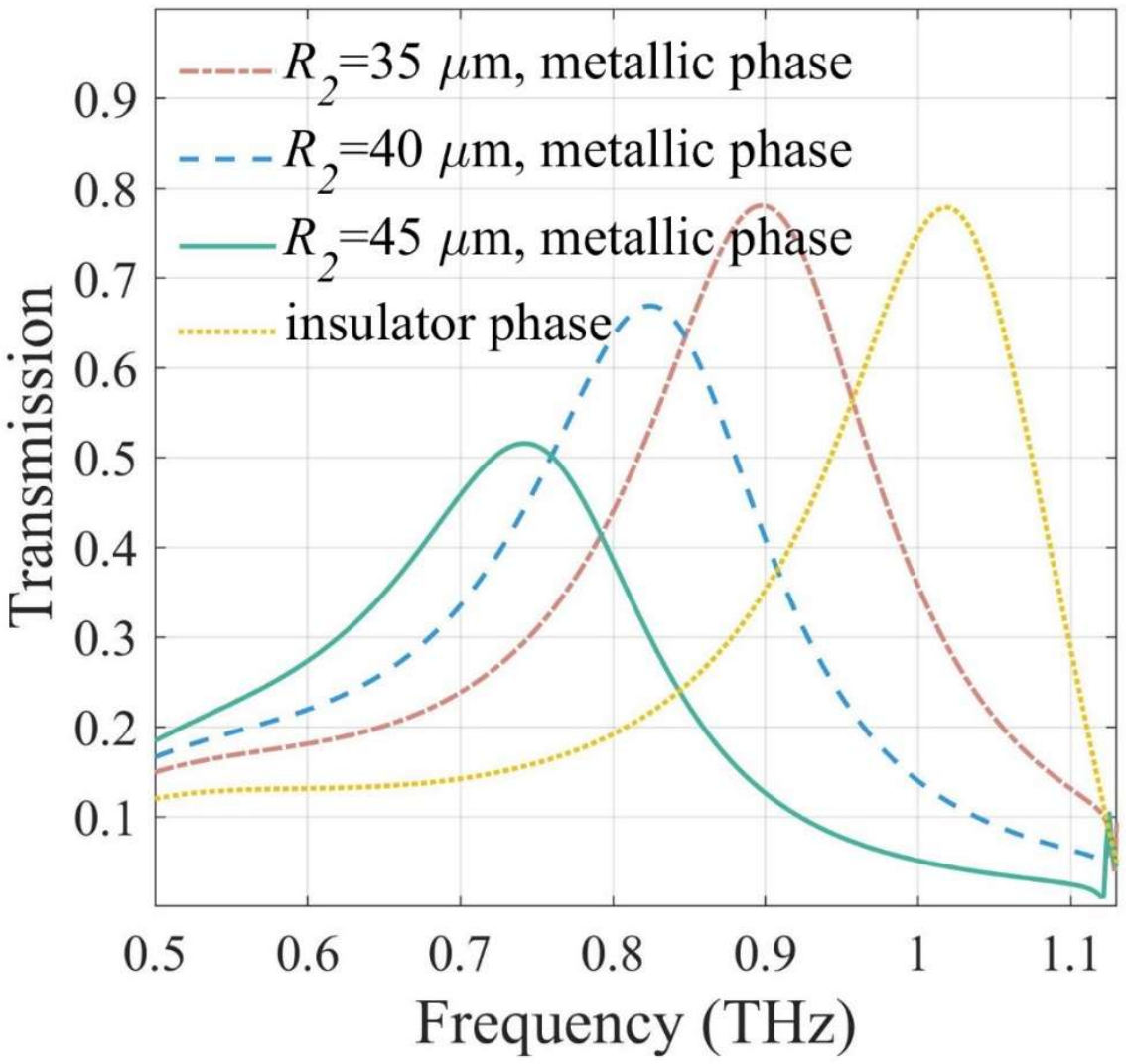

Fig. 6. Transmission resonance vs. the annular $VO_2$ disk's outer radius, with all the other geometrical parameters set to $t_m$=0.2 µm, $t_s$=75 µm, $R_3$=50 µm, $R_1$=20 µm, and $P$=150 µm. In the metallic phase, the transmission resonance redshifts as the size of the annular $VO_2$ disk increases. In the insulator phase, the $VO_2$ disk's radius has no effect on the transmission resonance since the hole shape is determined by the gold disk's size.

A narrower bandwidth is required for some applications, such as sensors. Hence, we investigate how the bandwidth of the transmission resonance might be reduced. To this end, we examine the performance of a simple array of holes in a 0.2 µm-thick gold plate with a radius of $R$ in a square lattice with a period of $P$. For simplicity, we remove the substrate and $VO_2$. As shown in Fig. 7(a), when the unit cell's size is fixed to $P$=150 µm, reducing the radius of the holes from $R$=60.0 to 22.5 µm blueshifts the transmission resonance from 1.86 to 2.58 THz with lower amplitude and narrower bandwidth. Full-width at half-maximum (FWHM) is 1.32 THz for $R$=60.0 µm while FWHM reduces to 0.07 THz as the size of hole decreases to $R$=22.5 µm. As seen in Fig. 7(b), while the radius of the hole remains constant at $R$=22.5 µm, increasing the size of unit cell from $P$=150 to 300 µm redshifts the resonance from 2.58 to 2.26 THz with narrower bandwidth and lower transmission levels. The FWHM decreases from 0.07 to 0.01 THz when the period increases from $P$=150 to 300 µm. In conclusion, the unit cell and hole size could be adjusted to achieve a narrow bandwidth resonance in the desired frequency range for sensing applications.

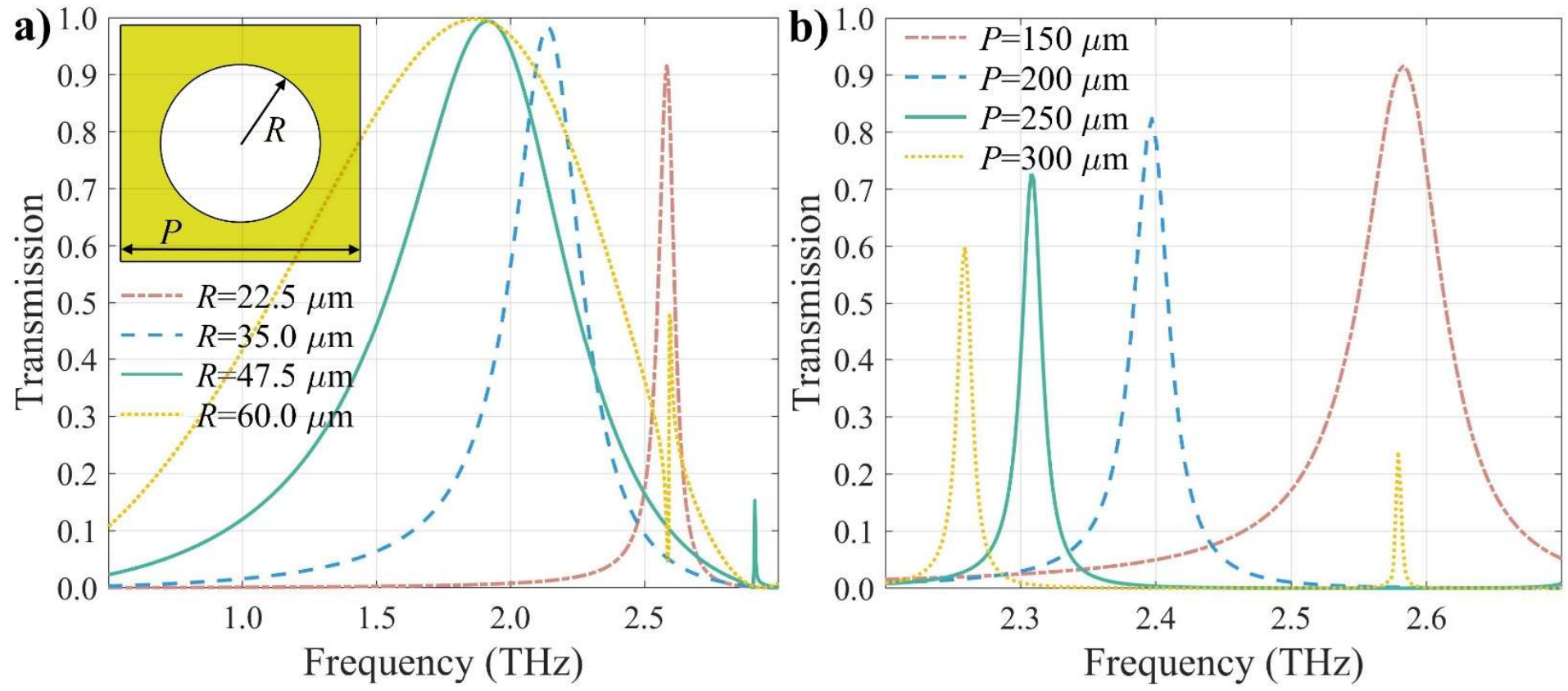

Fig. 7. Transmission resonance vs. a) the hole radius while $P$=150 µm and b) period of unit cell while $R$=22.5 µm. The size of the hole and unit cell can be adjusted to get the desired bandwidth at the required frequency.

### *2.2 Partial phase-transition of $VO_2$*

At room temperature, the $VO_2$ disks are in the insulator phase. Increasing the operating temperature of the metasurface leads to gradual nucleation and growth of metallic puddles within the dielectric $VO_2$. The properties of $VO_2$ in the intermediate state, i.e., when the metallic and insulator phases coexist in $VO_2$ can be calculated by effective medium theory. At about 85°C, the metallic puddles expand and form a continuous metallic film of $VO_2$ [42]. As the temperature increases, the conductivity of the $VO_2$ disk increases gradually from $2\times10^2$ to $2\times10^5$ S/m. The effect of partial phase-transition of annular $VO_2$ disk is shown in Fig. 8. As the conductivity of the annular $VO_2$ disk increase from $2\times10^2$ to $2\times10^5$ S/m, the transmission resonance at 1.02 THz shifts to higher frequencies while the transmission peak decreases to lower than 0.15. Meanwhile, a transmission resonance appears at lower frequencies of about 0.7 THz with a low transmission amplitude of 0.28. This transmission resonance gradually shifts to 0.82 THz with a higher transmission level of 0.67. As the conductivity increases to $\sigma=2\times10^2$, $6\times10^2$, $2\times10^3$, and $5\times10^3$ the transmission resonance blueshifts to the frequency of 1.02, 1.04, 1.09, and 1.12 THz while the transmission peak decreases to 0.78, 0.52, 0.22, and 0.15, respectively. Further increase of the conductivity to $\sigma=1\times10^4$, $5\times10^4$, $1\times10^5$, and $2\times10^5$ create a new transmission resonance at lower frequencies; this resonance blueshifts to the frequency of 0.70, 0.78, 0.81, and 0.82 THz while the transmission peak increases to 0.28, 0.48, 0.57, and 0.67, respectively. The conductivity level of $VO_2$ can be controlled thermally to modulate the frequency and amplitude of the transmission resonance of the designed metasurface.

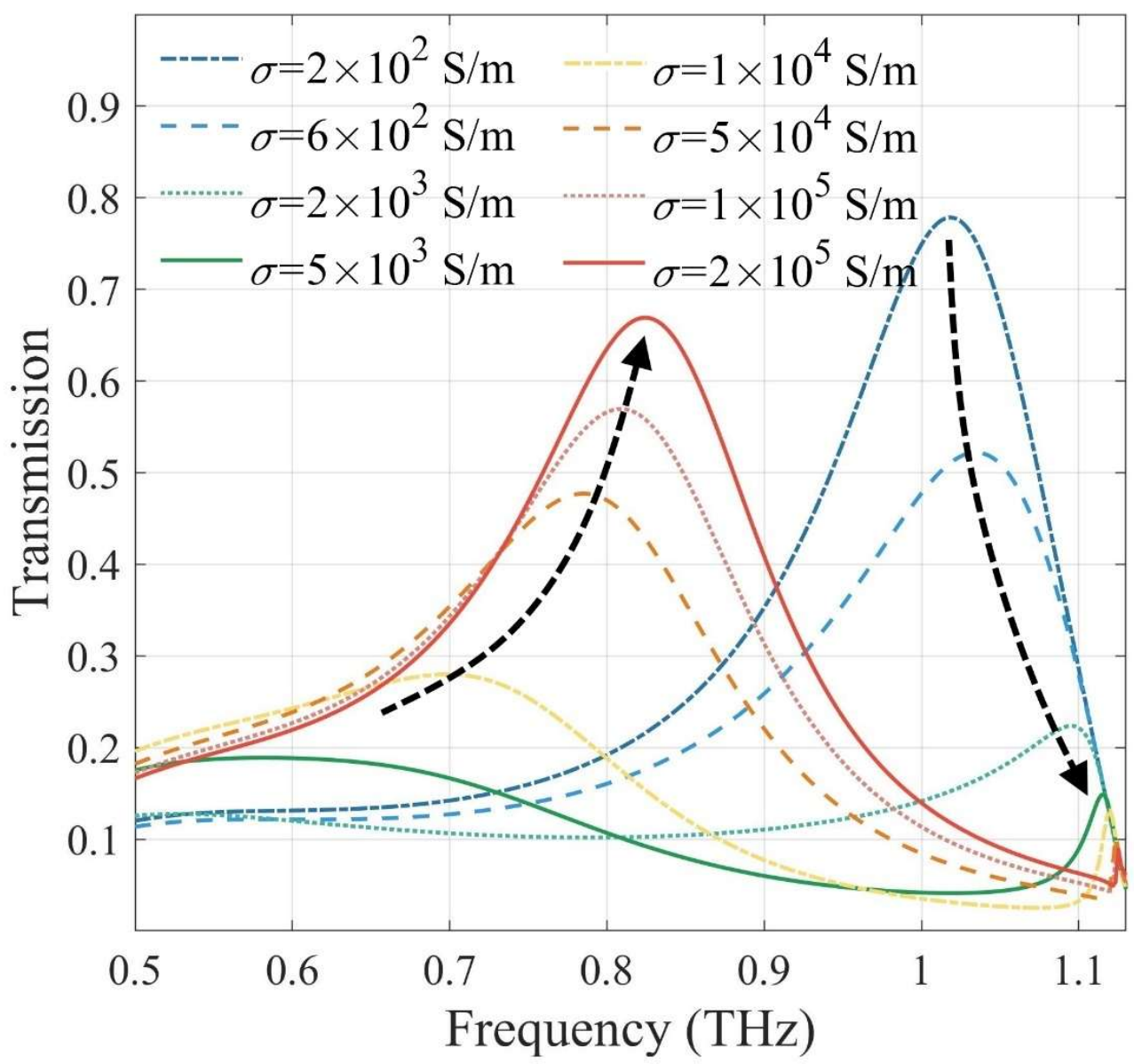

Fig. 8. Transmission resonance vs. the $VO_2$ conductivity. As the conductivity gradually increases, the transmission resonance at 1.02 THz blueshifts and vanishes while another transmission resonance appears at 0.82 THz.

## 3. Designing a blueshifting metasurface with EOT

In this section, we design a metasurface with a transmission resonance that blueshifts upon phase transition from the insulator to the metallic phase of the $VO_2$ disk. As shown in Fig. 9, the unit cell of the second metasurface is similar to the first design with a difference that, here, the inner disk is $VO_2$ which is surrounded by an annular gold disk. Similar to the previous design the thickness of the gold and $VO_2$ films is $t_m$=0.2 µm while the thickness of the $SiO_2$ substrate is $t_s$=75 µm. The radius of the $VO_2$ and gold disks and the radius of the hole are $R_1$=35 µm, $R_2$=45 µm, and $R_3$=50 µm, respectively. The period of the unit cell is $P$=150 µm. The transmission spectra of the second proposed metasurface are illustrated in Fig. 10. In the insulator phase, the transmission resonance occurs at the frequency of 0.71 THz with a peak amplitude of 0.66. As the operating temperature of the metasurface increases, the $VO_2$ transitions to the metallic phase. The transmission resonance blueshifts to 0.77 THz with a higher transmission peak of 0.84, in the metallic phase. The frequency shift, in this case, is merely 0.06 THz. The electric field distribution in the insulator and the metallic phases at the transmission resonances for the insulator and the metallic phases are shown in the inset of Fig. 10. The different mode of the aperture for the two phases of the $VO_2$ plays an important role in determining the transmission resonance characteristics.

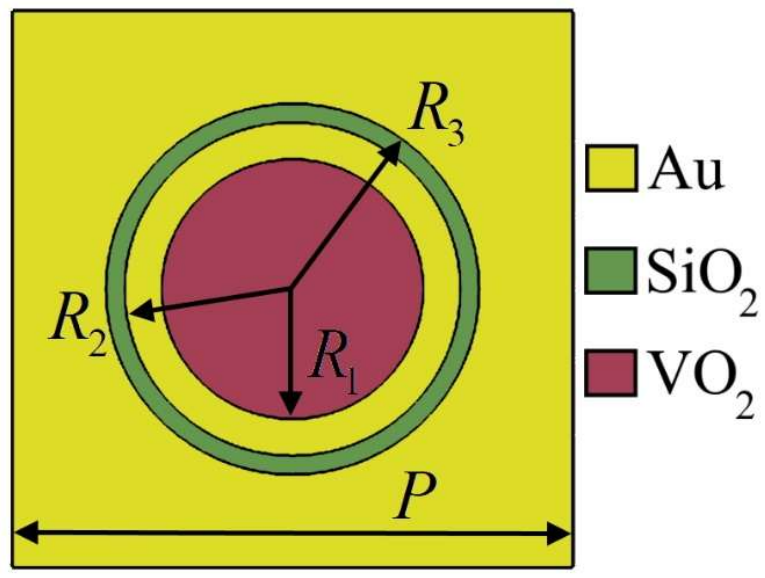

Fig. 9. Top view of the unit cell of the second structure. In this structure, upon phase-transition from the insulator to the metallic phase, the transmission resonance blueshifts to a higher frequency.

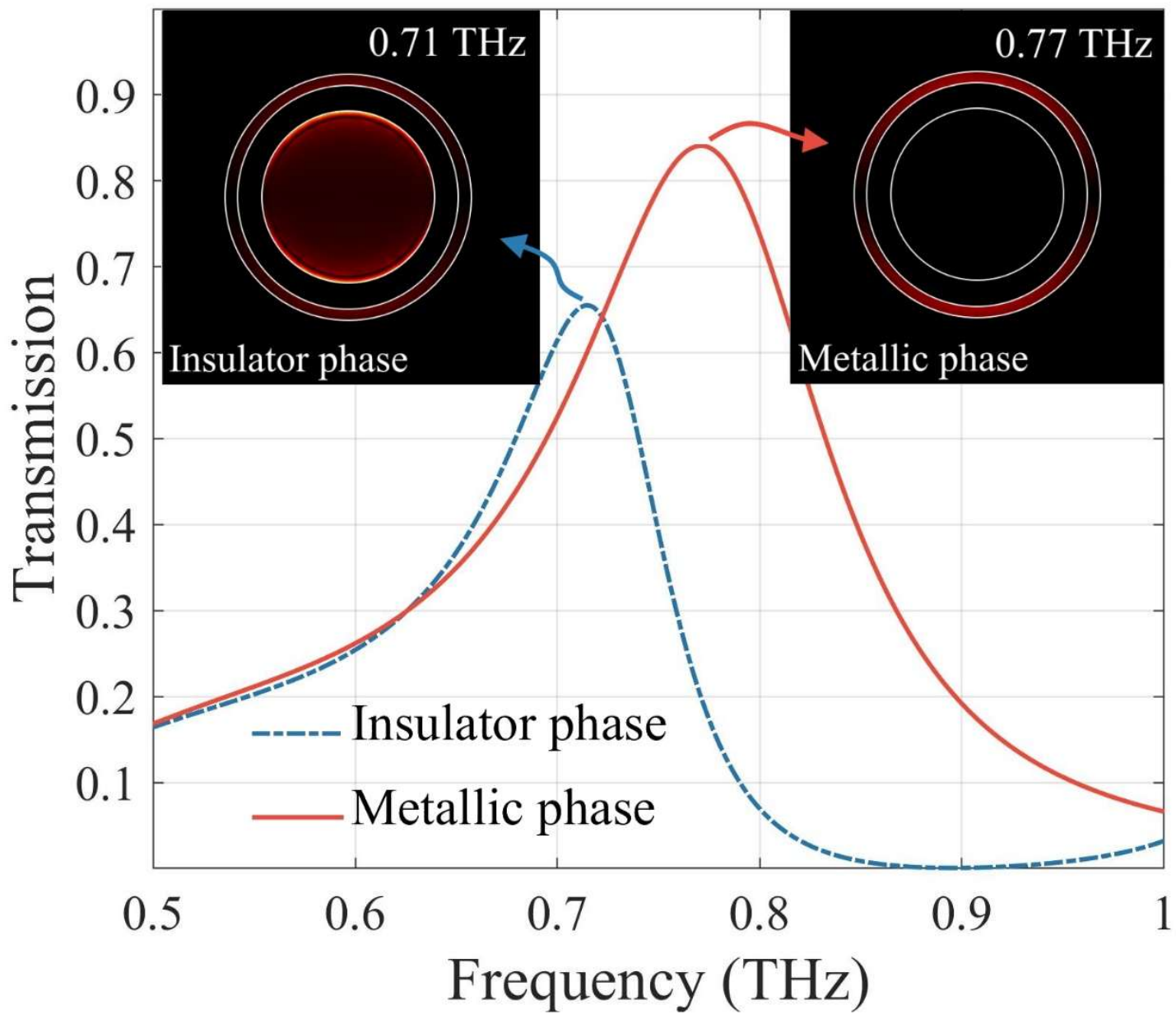

Fig. 10. The transmission spectra of the second proposed metasurface for the two phases of $VO_2$. The electric field distribution in the insulator and the metallic phases at the transmission resonance are shown.

As discussed in Introduction, surface plasmon modes on either side of the gold plate are coupled through holes. Therefore, the effective area of the holes and the supported modes inside the holes have a significant impact on the transmission resonance of the structure. The difference between our two designs is the position of the $VO_2$ and gold disks partially filling the holes. Because of this geometrical difference and the phase-transition of $VO_2$, distinct modes with different aperture area appear inside the holes. In the first design, the hole is an annular disk whose area changes upon phase-transition of $VO_2$ while the shape of the hole is not altered. The hole modes of the first design are shown in Fig. 3. However, in the second design, upon phase-transition from the insulator to the metallic phase, the shape of the hole changes noticeably. In the insulator phase, the hole is composed of an outer annular disk and an inner disk. But in the metallic phase, the hole is a single annular disk similar to the first design. In the insulator phase, the intensity of the electric field is higher in the inner disk as illustrated in the inset of Fig. 10. However, in the metallic phase, the electric field's intensity is concentrated in the outer annular disk. The differences in the shape and effective area of the holes are responsible for the redshifting and blueshifting responses of the first and second designs, respectively.

## 4. Comparison with previous studies

Finally, we compare the designed structures with other tunable EOT devices based on different mechanisms, including superconductor [15], graphene [16, 17], nematic liquid crystal [18], Schottky diode [19], piezoelectric [20], $VO_2$ [21], and $Ge_2Sb_2Te_5$ [22]. These approaches use thermal, electrical, magnetic, surface acoustic wave, or optical excitations to tune the transmission level or/and resonance frequency. Modulation depth is defined as $(A_1\text{-}A_2)\times100/(A_1+A_2)$, where $A_1$ and $A_2$ are the transmission peaks at the resonance frequency before and after applying the excitation. And $(f_1\text{-}f_2)/f_1$ is the normalized frequency shift, where $f_1$ and $f_2$ are the resonance frequencies before and after applying the excitation. We also compare the relative bandwidth of these designs before applying the excitation. Relative bandwidth is defined as $100\times\text{FWHM}/f_c$ where $f_c$ is the center frequency. In comparison to earlier devices operating at similar frequencies [15, 16, 18, 19, 22], our designs have a narrower bandwidth. It would not be constructive to compare the bandwidth with devices like [17, 20, 21], which operate at much higher frequencies. In comparison to earlier research, our first design has the largest redshift, as seen in Table 1. We also presented a second design that allows for resonance frequency blueshifting. Our designs aim for a higher frequency shift with a lower transmission amplitude change. Therefore, the modulation depth of the designed devices is lower than previous studies. The designed structures have the potential to be used as frequency modulators and reconfigurable filters.

Table 1. Comparison with other tunable EOT devices. For a fair comparison, we report the numerical results of other designs.

| Ref. | Mechanism | Excitation method | Resonance frequency (THz) | Normalized frequency shift | Modulation depth | Relative bandwidth |
|---|---|---|---|---|---|---|
| [15] | superconductor | thermal | 0.59 | - | 34.2% | 35.8% |
| [16] | graphene | electrical | 1.77 and 2.42 | - @ 1.77<br>0.17 (blueshift) @ 2.42 | 1.3% @ 1.77<br>90.6% @ 2.42 | 13.6% @ 1.77<br>18.9% @ 2.42 |
| [17] | graphene | electrical | 41.88 | - | 95.7% | 3.9% |
| [18] | nematic liquid crystal | magnetic | 0.193 | 0.03 (redshift) | 9.4% | 21.1% |
| [19] | Schottky diode | electrical | 0.75 | - | 35.7% | 25.3% |
| [20] | piezoelectric | surface acoustic wave | 205.1 | 0.009 (redshift) | - | 0.35% |
| [21] | $VO_2$ | thermal | 133.2 | - | 100% | - |
| [22] | $Ge_2Sb_2Te_5$ | optical | 0.85 | - | 88% | 18.8% |
| 1st design | $VO_2$ | thermal | 1.02 | 0.19 (redshift) | 7.6% | 18.0% |
| 2nd design | $VO_2$ | thermal | 0.71 | 0.08 (blueshift) | 12% | 17.1% |

## 5. Conclusion

In conclusion, we show that a structure with annular $VO_2$ and gold disks embedded in an array of holes in a gold film can support reconfigurable extraordinary terahertz transmission. The phase-transition of the $VO_2$ enables us to control the effective aperture area and mode of the aperture and, consequently, manipulate the frequency and amplitude of the transmission resonance. In the structure with holes partially filled with an inner gold disk and an outer annular $VO_2$ disk, the transmission redshift of 0.20 THz can be achieved upon the transition of the $VO_2$ from the insulator to the metallic phase. However, when an inner $VO_2$ disk and an outer annular gold disk fill the holes, the transmission resonance blueshifts by 0.06 THz. The designed metasurfaces can be used as reconfigurable filters or modulators in the terahertz frequencies.

**Disclosures.** The authors declare no conflicts of interest.

**Data availability.** Data underlying the results presented in this paper are not publicly available at this time but may be obtained from the authors upon reasonable request.